\documentclass{ws-mpla}

\renewcommand{\trimmarks}{ }
\begin{document}

\markboth{D. Ebert, R. N. Faustov \& V. O. Galkin}
{Relativistic description of the charmonium mass spectrum}

\title{\begin{flushright}\normalsize { HU-EP-05/10}\\
\vskip 1cm
\end{flushright}
RELATIVISTIC DESCRIPTION OF THE CHARMONIUM MASS SPECTRUM} 

\author{D. EBERT}

\address{Institut f\"ur Physik, Humboldt--Universit\"at zu Berlin,
Newtonstr. 15, D-12489  Berlin, Germany} 

\author{R. N. FAUSTOV and V. O. GALKIN}

\address{Institut f\"ur Physik, Humboldt--Universit\"at zu
  Berlin, Newtonstr. 15, D-12489  Berlin, Germany \\ 
and\\
Dorodnicyn Computing Centre, Russian Academy of Sciences, Vavilov Str. 40,
 119991 Moscow, Russia}

\maketitle

\begin{abstract}
The charmonium mass spectrum  is considered in the framework of the
constituent quark model with the relativistic treatment of the $c$
quark. The obtained masses are in good
agreement with the existing experimental data including the mass of
$\eta_c(2S)$. 

\end{abstract}

\ccode{PACS Nos.:  14.40.Gx, 12.39.Ki}

\vspace*{12pt}
\noindent Observation of the charmonium state $\eta'_c=\eta_c(2S)$
with the mass \cite{qwg,se}
\begin{equation}
  \label{eq:m}
  M_{\eta'_c}=3637.4\pm4.4 {\rm \ MeV (world\ average)}
\end{equation}
proved  to be in contradiction with most predictions of quark models
\cite{qwg,efg,bb}.  The hyperfine mass splitting in the $2S$ state was
found to be \cite{qwg,se} 
\begin{equation}
  \label{eq:2}
  \Delta M^{(2S)}_{hfs}=M_{\psi'}-M_{\eta_c'}=48.6\pm 4.4 {\rm\ MeV}.
\end{equation}
It was much smaller then the $1S$ splitting \cite{qwg,se}
\begin{equation}
  \label{eq:3}
  \Delta M^{(1S)}_{hfs}=M_{J/\psi}-M_{\eta_c}=115.1\pm 2.0 {\rm\ MeV},
\end{equation}
which was quite unexpected and surprising.

Different attempts to bring theory and experiment in accord were
undertaken. One of the most promising among them was the proposal to
take account of the open channel mixing \cite{qwg}.

Here we consider another possibility of describing charmonium with the
relativistic treatment of the $c$ quark. It is well-known (see
e.g. \cite{bb,bc}) that the $c$ quark is not heavy enough for a reliable
application of the nonrelativistic expansion since the $v^2/c^2$
contributions could exceed 20\%. This testifies a poor convergence of
the $v/c$ expansion. While describing the properties of heavy-light
mesons \cite{egf} we treated the light quarks in a completely relativistic
way. Now we shall apply this approach to recalculating the charmonium mass
spectrum in the framework of the previously developed relativistic
quark model.

  In this approach a meson is described by the wave
function of the bound quark-antiquark state, which satisfies the
quasipotential equation  of the Schr\"odinger type~\cite{efg}
\begin{equation}
\label{quas}
{\left(\frac{b^2(M)}{2\mu_{R}}-\frac{{\bf
p}^2}{2\mu_{R}}\right)\Psi_{M}({\bf p})} =\int\frac{d^3 q}{(2\pi)^3}
 V({\bf p,q};M)\Psi_{M}({\bf q}),
\end{equation}
where the relativistic reduced mass is
\begin{equation}
\mu_{R}=\frac{E_1E_2}{E_1+E_2}=\frac{M^4-(m^2_1-m^2_2)^2}{4M^3},
\end{equation}
and $E_1$, $E_2$ are given by
\begin{equation}
\label{ee}
E_1=\frac{M^2-m_2^2+m_1^2}{2M}, \quad E_2=\frac{M^2-m_1^2+m_2^2}{2M}.
\end{equation}
Here $M=E_1+E_2$ is the meson mass, $m_{1,2}$ are the quark masses,
and ${\bf p}$ is their relative momentum.  
In the center-of-mass system the relative momentum squared on mass shell 
reads
\begin{equation}
{b^2(M) }
=\frac{[M^2-(m_1+m_2)^2][M^2-(m_1-m_2)^2]}{4M^2}.
\end{equation}

The kernel 
$V({\bf p,q};M)$ in Eq.~(\ref{quas}) is the quasipotential operator of
the quark-antiquark interaction. It is constructed with the help of the
off-mass-shell scattering amplitude, projected onto the positive
energy states. 
Constructing the quasipotential of the quark-antiquark interaction, 
we have assumed that the effective
interaction is the sum of the usual one-gluon exchange term with the mixture
of long-range vector and scalar linear confining potentials, where
the vector confining potential
contains the Pauli interaction. The quasipotential is then defined by
  \begin{equation}
\label{qpot}
V({\bf p,q};M)=\bar{u}_1(p)\bar{u}_2(-p){\mathcal V}({\bf p}, {\bf
q};M)u_1(q)u_2(-q),
\end{equation}
with
$${\mathcal V}({\bf p},{\bf q};M)=\frac{4}{3}\alpha_sD_{ \mu\nu}({\bf
k})\gamma_1^{\mu}\gamma_2^{\nu}
+V^V_{\rm conf}({\bf k})\Gamma_1^{\mu}
\Gamma_{2;\mu}+V^S_{\rm conf}({\bf k}),$$
where $\alpha_S$ is the QCD coupling constant, $D_{\mu\nu}$ is the
gluon propagator in the Coulomb gauge
\begin{equation}
D^{00}({\bf k})=-\frac{4\pi}{{\bf k}^2}, \quad D^{ij}({\bf k})=
-\frac{4\pi}{k^2}\left(\delta^{ij}-\frac{k^ik^j}{{\bf k}^2}\right),
\quad D^{0i}=D^{i0}=0,
\end{equation}
and ${\bf k=p-q}$; $\gamma_{\mu}$ and $u(p)$ are 
the Dirac matrices and spinors
\begin{equation}
\label{spinor}
u^\lambda({p})=\sqrt{\frac{\epsilon(p)+m}{2\epsilon(p)}}
\left(
\begin{array}{c}1\cr {\displaystyle\frac{\bm{\sigma}
      {\bf  p}}{\epsilon(p)+m}}
\end{array}\right)\chi^\lambda,
\end{equation}
with $\epsilon(p)=\sqrt{p^2+m^2}$.
The effective long-range vector vertex is
given by
\begin{equation}
\label{kappa}
\Gamma_{\mu}({\bf k})=\gamma_{\mu}+
\frac{i\kappa}{2m}\sigma_{\mu\nu}k^{\nu},
\end{equation}
where $\kappa$ is the Pauli interaction constant characterizing the
anomalous chromomagnetic moment of quarks. Vector and
scalar confining potentials in the nonrelativistic limit reduce to
\begin{eqnarray}
\label{vlin}
V_V(r)&=&(1-\varepsilon)Ar+B,\nonumber\\ 
V_S(r)& =&\varepsilon Ar,
\end{eqnarray}
reproducing 
\begin{equation}
\label{nr}
V_{\rm conf}(r)=V_S(r)+V_V(r)=Ar+B,
\end{equation}
where $\varepsilon$ is the mixing coefficient. 

The $c$ quark mass $m_c=1.55$ GeV and
the parameters of the linear potential $A=0.18$ GeV$^2$ and $B=-0.16$ GeV
have usual values of quark models.  The value of the mixing
coefficient of vector and scalar confining potentials $\varepsilon=-1$
has been determined from the consideration of charmonium radiative
decays \cite{efg}. 
Finally, the universal Pauli interaction constant $\kappa=-1$ has been
fixed from the analysis of the fine splitting of heavy quarkonia ${
}^3P_J$- states \cite{efg}. 

In oder to simplify the relativistic $q\bar q$ potential we make the
following replacement in the Dirac spinors:
\begin{equation}
  \label{eq:sub}
  \epsilon(p)=\sqrt{m^2+{\bf p}^2} \to E=M/2
\end{equation}
(see the discussion of this point in \cite{egf}).

As a result the spin-independent part of the potential with the
account of the one-loop radiative corrections reads as
\begin{eqnarray}
\label{sipot}
V_{\rm SI}(r)&=&
\left(\frac{E+m}{2E}\right)^2\left[-\frac43\frac{\bar
    \alpha_V(\mu^2)}{r} +Ar+B\right] -\frac43 
\frac{\beta_0 \alpha_s^2(\mu^2)}{2\pi}\frac{\ln(\mu r)}{r} \cr
&&
+\frac1{4E^2}\left[{\bf p}^2,-\frac43\frac{\bar \alpha_V(\mu^2)}{r} +Ar+B\right]_+
\cr
&& +\frac14
\Delta\Biggl( \frac1{E^2}\left[-\frac43\frac{\bar 
\alpha_V(\mu^2)}{r}
 +(1-\varepsilon)\left(1+2\frac{E+m}{2m}\kappa\right)Ar\right]\cr
&&
 -\frac1{m^2}\frac43\frac{\beta_0\alpha_s^2(\mu^2)}{2\pi}
\frac{\ln(\mu r)}{r}\Biggr)
+\frac{1}{2E^2}\Biggl(\left\{-\frac43\frac{\bar\alpha_V}{r}
\left[{\bf p}^2+\frac{({\bf p\cdot r})^2}{r^2}\right]\right\}_W\cr
&& +(1-2\varepsilon)
\left\{Ar\left[{\bf p}^2
-\frac{({\bf p\cdot r})^2}{r^2}\right]\right\}_W+
3B{\bf p}^2\Biggr)\cr
&&-\frac1{2m^2}\frac43\frac{\beta_0\alpha_s^2(\mu^2)}{2\pi}\left\{{\bf p}^2
  \frac{\ln(\mu r)}{r} +\frac{({\bf p\cdot 
r})^2}{r^2}\left(\frac{\ln(\mu r)}{r}-\frac1r\right)\right\}_W 
+O\!\left(\frac{1}{m^4}\right),\cr&&
\end{eqnarray}
where $\{\dots\}_W$ denotes the Weyl ordering of operators and 
\begin{eqnarray*}
&&\bar\alpha_V(\mu^2)=\alpha_s(\mu^2)\left[1+\left(\frac{a_1}{4}
+\frac{\gamma_E\beta_0}{2}\right)\frac{\alpha_s(\mu^2)}{\pi}\right],\cr
&&a_1=\frac{31}{3}-\frac{10}{9}n_f,\quad
\beta_0=11-\frac23n_f,\quad n_f=3,\cr
&&
\alpha_s(\mu^2)=\frac{4\pi}{\beta_0\ln(\mu^2/\Lambda^2)},
\qquad \mu=E, \qquad \Lambda=169\ {\rm MeV}.
\end{eqnarray*}
The spin-dependent part of the potential has the form:
\begin{equation}
\label{vsd}
V_{\rm SD}(r)
= a\ {\bf L}\cdot{\bf S}+b\left[\frac{3}{r^2}({\bf S}_1\cdot {\bf r})
({\bf S}_2\cdot {\bf r})-({\bf S}_1\cdot {\bf S}_2)\right] +c\ {\bf
S}_1\cdot {\bf S}_2, 
\end{equation}
\begin{eqnarray}
\label{a}
a&=& \frac{1}{2E^2}\Biggl\{\frac{4\alpha_s(\mu^2)}{r^3}\Biggl(1+\frac{E^2}{m^2}
\frac{\alpha_s(\mu^2)}{\pi}\Biggl[\frac{1}{18}n_f-\frac{1}{36}+\gamma_E\left(
\frac{\beta_0}{2}-2\right)\cr
&&+\frac{\beta_0}{2}\ln\frac{\mu}{m}+\left(\frac{\beta_0}{2}-2\right)\ln(mr)\Biggr]\Biggr)
-\left(\frac{E}{m}-\frac{E-m}{m}\varepsilon\right)\frac{A}{r}\cr
&&+4(1+\kappa)\frac{E+m}{2m}(1-\varepsilon)\frac{A}{r}\Biggr\}\\
\label{b}
b&=& \frac{1}{3E^2}\Biggl\{\frac{4\alpha_s(\mu^2)}{r^3}\Biggl(1+\frac{E^2}{m^2}
\frac{\alpha_s(\mu^2)}{\pi}\Biggl[\frac{1}{6}n_f+\frac{25}{12}+\gamma_E\left(
\frac{\beta_0}{2}-3\right)+\frac{\beta_0}{2}\ln\frac{\mu}{m}\cr
&&+\left(\frac{\beta_0}{2}-3\right)\ln(mr)\Biggr]\Biggr)
+\left(\frac{E-m}{2m}-(1+\kappa)\frac{E+m}{2m}
\right)^2(1-\varepsilon)\frac{A}{r}\Biggr\}\\ 
\label{c}
c&=& \frac{4}{3E^2}\Biggl\{\frac{8\pi\alpha_s(\mu^2)}{3}\Biggl(\left[1+\frac{E^2}{m^2}
\frac{\alpha_s(\mu^2)}{\pi}\left(\frac{23}{12}-\frac{5}{18}n_f-\frac34\ln2\right)
\right]\delta^3(r)\cr
&&+\frac{E^2}{m^2}\frac{\alpha_s(\mu^2)}{\pi}\Biggl[-
\frac{\beta_0}{8\pi}\nabla^2\left(\frac{\ln({\mu}/{m})}{r}\right)
+\frac{1}{\pi}\left(\frac{1}{12}n_f-\frac{1}{16}\right)\cr
&&\times\nabla^2\left(
\frac{\ln(mr)+\gamma_E}{r}\right)\Biggr]\Biggr)+\left(
\frac{E-m}{2m}-(1+\kappa)\frac{E+m}{2m} \right)^2(1-\varepsilon)\frac{A}{r}\Biggr\}.
\end{eqnarray}
In Eqs. (\ref{sipot})--(\ref{c}) the scale $\mu=E$ is state
dependent. The one-loop corrections are known only for the $1/m^2$
contributions in the above expressions and will be treated perturbatively.

In the initial approximation the quasipotential equation (\ref{quas})
is solved numerically with the following potential
\begin{equation}
  \label{eq:vo}
  V_0(r)=\left(\frac{E+m}{2E}\right)^2\left[-\frac43\frac{\bar
    \alpha_V(\mu^2)}{r} +Ar+B\right].
\end{equation}
The remaining contributions in $V_{SI}$ and $V_{SD}$ are treated
perturbatively. It is important to mention that since the scale is
$\mu=E$, the QCD coupling constant $\alpha_s$ (and also $\alpha_V$)
take different values dependent on quantum numbers $n$ and $L$ (see
Table~1). The obtained mass spectrum is presented in Table~1, cf \cite{bb}.
\begin{table}[htbp]
\tbl{Charmonium mass spectrum (in GeV). }
{\begin{tabular}{cclcc|c}
\hline
State  & & {Experiment}& \multicolumn{2}{c|}{Theory} & \\
$n\,{}^{2S+1}\! L_J$& Particle &{PDG \cite{pdg}} &  old
\cite{efg}&new & $\alpha$\\
\hline
$1\,{}^1\! S_0$& $\eta_c$  & 2.9796 & 2.979& 2.978&$\alpha_s=0.315$\\
$1\,{}^3\! S_1$& $J/\Psi$  & 3.09692 & 3.096 &3.097&$\bar\alpha_V=0.452$\\
&&&&&\\
$1\,{}^3\! P_0$& $\chi_{c0}$ & 3.41519&   3.424 &3.423\\
$1\,{}^3\! P_1$& $\chi_{c1}$  & 3.51059  & 3.510 &3.509&$\alpha_s=0.298$\\
$1\,{}^3\! P_2$& $\chi_{c2}$  & 3.55626  & 3.556 &3.556&$\bar\alpha_V=0.421$ \\
$1\,{}^1\! P_1$& $h_c$&3.5244 \cite{se} &   3.526& 3.525\\
&&&&&\\
$2\,{}^1\! S_0$& $\eta_c'$  & 3.6374 \cite{se} & 3.588& 3.633&$\alpha_s=0.291$\\
$2\,{}^3\! S_1$& $\Psi'$     & 3.68609   & 3.686& 3.684&$\bar\alpha_V=0.408$\\ 
&&&&&\\
$1\,{}^3\! D_1$&   & 3.7700  & 3.798&3.795\\
$1\,{}^3\! D_2$&   &  & 3.813& 3.810&$\alpha_s=0.288$\\
$1\,{}^3\! D_3$&   &  & 3.815&3.816&$\bar\alpha_V=0.403$\\
$1\,{}^1\! D_2$& & & 3.811&3.810\\
\hline
\end{tabular}} 
\end{table} 

Note that all spin-splittings (including the hyperfine ones) are well
reproduced and the new prediction for 
the mass of $\eta_c(2S)$, $M^{th}=3633$~MeV, is close to the experimental value
(\ref{eq:m}). The calculated center-of-gravity mass of the $1^3P_J$ states
$M^{th}_{cog}=3525$~MeV is equal to the predicted of the $1^1P_1$
state $M^{th}_{h_c}=3525$~MeV up to terms of order 0.1~MeV and is
close to the PDG value 3525.36~MeV. This signifies that the
relativistic and scale-dependent 
effects are essential for the consistent description of the
charmonium. The latter statement agrees with the conclusions of
Ref.\cite{bb}. 

The hyperfine mass splitting is intimately connected with the leptonic
decay rate of heavy quarkonium \cite{bb,mpl}. For the $1^3S_1$ state
($J/\psi$) this decay rate is the same (5.4 keV) as in our paper
\cite{mpl} since the factor in front of the square brackets in
Eq.~(\ref{eq:vo}) is almost unity. For the $2^3S_1$ state ($\psi'$) we
recalculated its leptonic decay rate using the new wave function and
obtained $\Gamma_{ee}^{th}=2.0$~keV. Thus our prediction is in accord
with the experimental value $\Gamma_{ee}^{exp}=2.12\pm0.12$~keV
\cite{pdg}. 

The authors are grateful to A. Badalian, M. M\"uller-Preussker and
V. Savrin for support and discussions.  Two of us (R.N.F. and V.O.G.) 
were supported in part by the {\it Deutsche
Forschungsgemeinschaft} under contract Eb 139/2-3. 



\section*{References}
\begin{thebibliography}{00}
\bibitem{qwg} Quarkonium Working Group, ``Heavy Quarkonium Physics'',
  CERN Yellow Report, hep-ph/0412158.
\bibitem{se} K.K. Seth, hep-ex/0501022.
\bibitem{efg} D. Ebert, R.N. Faustov and V.O. Galkin, {\it
    Phys. Rev. D} {\bf 67}, 014027 (2003).
\bibitem{bb} A.M. Badalian and B.L.G. Bakker, {\it Phys. Rev. D} {\bf
    67}, 071901 (2003).
\bibitem{bc} A.E. Bondar and V.L. Chernyak, hep-ph/0412335.
\bibitem{egf} D. Ebert, V.O. Galkin and R.N. Faustov, {\it
    Phys. Rev. D} {\bf 57}, 5663 (1998).
\bibitem{mpl} D. Ebert, R.N. Faustov  and V.O. Galkin,
  {\it Mod. Phys. Lett. A} {\bf 18}, 1597 (2003).
\bibitem{pdg} Particle Data Group, S. Eidelman et al., {\it
    Phys. Lett. B} {\bf 592}, 1 (2004).


\end{thebibliography}
\end{document}